\documentclass[aps,prd,floatfix,nofootinbib]{revtex4} 
\usepackage{graphics}
\usepackage{graphicx}

\usepackage{latexsym}
\usepackage[cp866]{inputenc}
\usepackage[english]{babel}
\input epsf

\begin{document}

\title{\boldmath Toward an estimate of the amplitude $X(3872)\to\pi^0\chi_{c1}(1P)$}
\author{N. N. Achasov\,\footnote{achasov@math.nsc.ru}
and G. N. Shestakov\,\footnote{shestako@math.nsc.ru}}
\affiliation{\vspace{0.2cm} Laboratory of Theoretical Physics, S. L.
Sobolev Institute for Mathematics, 630090, Novosibirsk, Russia}


\begin{abstract}

The well-known model of the triangle diagrams with $D^*\bar DD^*$
and $\bar D^*D\bar D^*$ mesons in the loops is compared with the
modern data on the amplitude of the $X(3872)\to\pi^0\chi_{c1}(1P)$
decay. Considering the $X(3872)$ object as a $\chi_{c1}(2P)$
charmonium state, we introduce a parameter $\xi$ characterizing the
scale of the isotopic symmetry violation in this decay and find a
lower limit of $\xi\simeq0.0916$. The model incorporates the only
fitted parameter associated with the form factor. We analyze in
detail the influence of the form factor on the amplitude
$X(3872)\to\pi^0 \chi_{c1}(1P)$ and on the parameter $\xi$. As the
suppression of the amplitude by the form factor increases, $\xi$
increases. Because the $X(3872)$ resonance is located practically at
the threshold of the $D^0\bar D^{*0}$ channel, the amplitude of
$X(3872) \to\pi^0\chi_{c1}(1P)$ turns out to be proportional to
$\sqrt{m_d -m_u}$. Using the estimating values for the coupling
constants $g_{XD\bar D^*}$, $g_{\chi_{ c1}D\bar D^*}$, and
$g_{D^{*0}D^{*0}\pi^0}$, we show that the model of the triangle loop
diagrams is in reasonable agreement with the available data. Apart
from the difference in the masses of neutral and charged charmed
mesons, any additional exotic sources of isospin violation in
$X(3872)\to\pi^0\chi_{c1}(1P)$ (such as a significant difference
between the coupling constants $g_{XD^0\bar D^{*0}}$ and $g_{XD^+
D^{*-}}$) are not required to interpret the data. This indirectly
confirms the isotopic neutrality of the $X(3872)$, which is
naturally realized for the $c\bar c$ state $\chi_{c1}(2P)$.

\end{abstract}

\maketitle

\section{Introduction}


The state $X(3872)$ or $\chi_{c1}(3872)$ \cite{PDG23} was observed
for the first time by the Belle Collaboration in 2003 in the process
$B^\pm\to(X(3872)\to\pi^+\pi^-J/\psi)K^\pm$ \cite{Cho03}. Then it
was observed in many other experiments in other processes and decay
channels \cite{PDG23,Kop23}. The $X(3872)$ is a very narrow
resonance. Its visible width depends on the decay channel. In the
$\pi^+\pi^-J/\psi$ channel, the width of the $X(3872)$ peak is
approximately of 1 MeV \cite{PDG23,Cho11,Aai20} and in the $(D^{*0}
\bar D^0+\bar D^{*0}D^0)\to D^0\bar D^0\pi^0$ channel, it is of
about 2--5 MeV \cite{PDG23,Gok06,Aus10,Hir23,Abl23,Tan23}. Its mass
coincides practically with the $D^{*0}\bar D^0$ threshold
\cite{PDG23}. The $X(3872)$ has the quantum numbers $I^G(J^{PC})
=0^+(1^{++})$ \cite{PDG23,Aub05,Cho11,Aai13,Aai15}. In addition to
decays into $\pi^+\pi^-J/\psi$ \cite{Cho03,Cho11,Aai13, Abl14,Aai20}
and $D^0\bar D^0\pi^0$ \cite{Gok06,Aus10,Hir23, Abl23,Tan23}, the
$X(3872)$ also decays into $\omega J/\psi$ \cite{Abe05,Amo10,Abl19},
$\gamma J/\psi$ \cite{Abe05,Aub09,Bha11, Aai14,Abl20}, $\gamma\psi
(2S)$ \cite{Aub09,Bha11,Aai14}, and $\pi^0\chi_{c1}(1P)$ \cite{Ab19,
Bh19}. The $X(3872)$ became the first candidate for exotic
charmoniumlike states, and many hypotheses have been put forward
about its nature; see Refs.
\cite{PDG23,Cho03,Kop23,Cho11,Aai20,Gok06,Aus10,Hir23,Abe05,Abl14,
Abl20,Abl23,Tan23,Aub05,Aai13,Aai15,Amo10,Abl19,Aub09,Bha11,Aai14,
Ab19,Bh19,MR23a,Sw04,Zh14,Ma05,AR14,AR15,AR16,AKS22,Ka05,TT13,Su05}
and references herein. For example, the $X(3872)$ is interpreted as
a hadronic $D\bar D^*$ molecule \cite{Sw04,Zh14}, a compact
tetraquark state \cite{Ma05}, a conventional charmonium state
$\chi_{c1}(2P)$ \cite{AR14,AR15,AR16, AKS22}, a mixture of a
molecule, and an excited charmonium state \cite{Ka05,TT13,Su05},
etc. So far, none of these explanations have become generally
accepted. But there is hope that new, more and more accurate
experiments will allow us to make a definite choice between the
different interpretations.

Of great interest are the $X(3872)$ decays that violate isospin:
$X(3872)\to\pi^+\pi^-J/\psi$, $X(3872)\to\pi^0\chi_{c1}(1P)$, and
$X(3872)\to\pi^0\pi^+\pi^-$ \cite{Abe05,Amo10,Abl19,Ab19,Bh19,To04,
Su05,Me07,Os09,Os10,Te10,KL10,Li12,Ac19,Zh19,Wu21,Me21,Aa23,Wa23,
DV08,FM08,Me15,AS19,Yi21}. In what follows, we will discuss the
$X(3872)\to\pi^0\chi_{c1}(1P)$ decay. Even before the appearance of
the BESIII \cite{Ab19} and Belle \cite{Bh19} data (see also
\cite{PDG23}), a number of model predictions were made for it
\cite{DV08,FM08,Me15}. Then this decay was studied in the works
\cite{Zh19,Wu21,Wa23}. In Ref. \cite{DV08}, under the assumption
that $X(3872)$ is a conventional $c\bar c$ state and that $\pi^0$ is
produced in its decay via two-gluon mechanism, the value of $\simeq
0.06$ keV was obtained for the width $\Gamma(X(3872)\to\pi^0\chi_{
c1}(1P))$, which is several orders of magnitude less than what
follows from the experiment \cite{Ab19}. In Ref. \cite{FM08}, the
$X(3872)$ was considered as a loosely bound state of neutral charmed
mesons $D^0\bar D^{*0}+\bar D^0D^{*0}$. If the decay of such a
molecular quarkonium into $\pi^0\chi_{c1}(1P)$ results from the
neutral charmed meson loop mechanism, then, according to the
estimate \cite{Me15}, $\Gamma(X(3872)\to\pi^0 \chi_{c1}(1P))$ turns
out to be greater than the total $X(3872)$ width. To avoid
contradictions with experiment, it was proposed \cite{Me15} to take
into account the coupling of the $X(3872)$ to charged charmed mesons
$D^+D^{*-}+D^- D^{*+}$. In this case, the contributions of the
triangle loops with neutral and charged $D^{(*)}$ mesons should
partially compensate each other in the transition amplitude $X(3872
)\to\pi^0\chi_{c1}(1P)$ \cite{Me15}, which is completely natural for
the $X(3872)$ state with isospin $I=0$. In Ref. \cite{Zh19}, to
describe the $X(3872)$, a scheme was used in which $D\bar D^*$ pairs
were considered as the dominant components in its wave function, and
it was obtained that $\Gamma( X(3872)\to \pi^0\chi_{c1}(1P))$ is an
order of magnitude smaller than $\Gamma(X(3872)\to\pi^+\pi^-J/\psi
)$. In Ref. \cite{Wu21}, the molecular scenario  for the $X(3872)$
was considered. It was assumed that the strong isospin violation in
the decays $X(3872)\to\pi^+\pi^- J/\psi$, $X(3872)\to\pi^+\pi^-\pi^0
J/\psi$, and $X(3872)\to\pi^0 \chi_{cJ}(1P)$ comes from the
different coupling strengths of the $X(3872)$ to its charged
$D^+D^{*-}$ and neutral $D^0\bar D^{*0}$ components as well as
through the interference between the charged and neutral meson
loops. In Ref. \cite{Wu21}, the nonstandard normalizations were used
for $\Gamma(X(3872)\to\pi^+\pi^-J/\psi)$ and $\Gamma(X(3872)\to
\pi^+\pi^-\pi^0J/\psi)$ (see Ref. \cite{FN1}), and therefore, the
agreement with experiment obtained for the ratio $\Gamma(X(3872)
\to\pi^0 \chi_{c1}(1P))/\Gamma(X(3872)\to\pi^+\pi^-J/\psi)$ is
doubtful. In Ref. \cite{Wa23}, the $X(3872)$ was considered as a
tetraquark state with the $I=0$ and 1 isospin components, and its
decays were analyzed via the QCD sum rules. In so doing, for
$\Gamma(X(3872)\to\pi^0\chi_{c1}(1P))$ the value of $\approx0.0016$
MeV was obtained, which is approximately 20 times smaller in
comparison with the experimental estimate \cite {PDG23}.

In the present work, we consider the $X(3872)$ meson as a
$\chi_{c1}(2P)$ charmonium state, which has the equal coupling
constants with the $D^0\bar D^{*0}$ and $ D^+ D^{*-} $ channels
owing to the isotopic symmetry. Section II collects the available
data on the $X(3872)\to\pi^0\chi_{c1}(1P)$ decay. In Sec. III, we
calculate the transition amplitude $X(3872)\to\pi^0\chi_{c1}(1P)$
corresponding to the simplest $D^*\bar DD^*+c.c.$ loop mechanism
\cite{Me15,Wu21}, we pay attention to details that were not
previously discussed, and introduce the parameter $\xi$
characterizing the natural scale of isospin violation for the
process under consideration. In Sec. IV, we analyze in detail the
influence of the form factor on the magnitude of the amplitude
$X(3872)\to\pi^0\chi_{c1}(1P)$ and on the parameter $\xi$. Using the
evaluating values for coupling constants $g_{XD\bar D^*}$,
$g_{\chi_{c1}D\bar D^*}$, and $g_{D^{*0}D^{*0}\pi^0}$, we show that
the model of charmed meson loops explains the data on the absolute
value of the amplitude of the $X(3872)\to \pi^0\chi_ {c1}(1P)$ decay
by a quite naturally way. Our conclusions from the presented
analysis are given in Sec. V, together with a short comment
regarding the molecular model of the $X(3872)$ state.


\section{Data on the \boldmath{$X(3872)\to\pi^0\chi_{\lowercase{c}1}(1P)$} decay}


Let us write the transition amplitude $X(3872)\to\pi^0\chi_{c1}(1P)$
in the form,
\begin{eqnarray}\label{Eq1}
\mathcal{M}(X(3872)\to\pi^0\chi_{c1}(1P);s)\equiv\mathcal{M}_{\pi^0}(s)=
\varepsilon^{\mu\nu\lambda\kappa}\epsilon^X_\mu(p_1)\epsilon^{
*\chi_{c1}}_\nu(p_2)p_{1\lambda}p_{2\kappa}G_{\pi^0}(s),\end{eqnarray}
where $\epsilon^X(p_1)$ and $\epsilon^{*\chi_{c1}}(p_2)$ are the
polarization four-vectors of the $X(3872)$ and $\chi_{c1}(1P)$
mesons, respectively (helicity indices omitted), $p_1$, $p_2$ and
$p_3=p_1-p_2$ are the four-momenta of $X(3872)$, $\chi_{c1}(1P)$ and
$\pi^0$, respectively, $s=(p_2+p_3)^2$ is the squared invariant mass
of the $\pi^0\chi_{c1}(1P)$ system or of the virtual $X(3872)$
state, and $G_{\pi ^0}(s)$ is the invariant amplitude. The
energy-dependent width of the $X(3872)\to\pi^0\chi_{c1}(1P)$ decay
in the rest frame of $X(3872)$ is expressed in terms of
$G_{\pi^0}(s)$ as follows :
\begin{eqnarray}\label{Eq2}
\Gamma(X(3872)\to\pi^0\chi_{c1}(1P);s)=\frac{|G_{\pi^0}(s)|^2}{12\pi}\,
|\vec{p}_3|^3,\end{eqnarray} where $|\vec{p}_3|=\sqrt{s^2
-2s(m^2_{\chi_{c1}}+m^2_{\pi^0})+(m^2_{\chi_{c1}}-m^2_{\pi^0})^2}
/(2\sqrt{s})$. The following information is available about the
decay of $X(3872)\to\pi^0\chi_{c1}(1P)$. The BESIII Collaboration
\cite{Ab19} observed this decay and determined the value of the
ratio,
\begin{eqnarray}\label{Eq3}
\frac{\mathcal{B}(X(3872)\to\pi^0\chi_{c1}(1P))}{\mathcal{B}(X(3872)\to
\pi^+\pi^-J/\psi)}=0.88^{+033}_{-027}\pm0.10.\end{eqnarray} The
Belle Collaboration \cite{Bh19} set an upper limit for this ratio,
\begin{eqnarray}\label{Eq4}
\frac{\mathcal{B}(X(3872)\to\pi^0\chi_{c1}(1P))}{\mathcal{B}(X(3872)\to
\pi^+\pi^-J/\psi)}<0.97\end{eqnarray} at the 90\% confidence level.
The Particle Data Group (PDG)\cite{PDG23} gives for the branching
fraction of $X(3872)\to\pi^0\chi_{c1}(1P)$ the following value:
\begin{eqnarray}\label{Eq5}
\mathcal{B}(X(3872)\to\pi^0\chi_{c1}(1P))=(3.4\pm1.6)\%,\end{eqnarray}
and also gives a constraint $\mathcal{B}
(X(3872)\to\pi^0\chi_{c1}(1P))<4\%$ based on the Belle data.
Moreover, according to the analysis presented in Ref. \cite{LY19},
$\mathcal{B}(X(3872)\to\pi^0\chi_{c1}(1P))=(3.6^{+2.2}_{-1.6})\%$.

Using Eqs. (\ref{Eq2}), (\ref{Eq5}) and the value of the $X(3872)$
total decay width presented by the PDG \cite{PDG23},
$\Gamma^{\scriptsize\mbox{tot}}_X=(1.19\pm0.21)$ MeV, we obtain the
following approximate estimates for the absolute decay width of
$X(3872)\to\pi^0\chi_{c1}(1P)$ and for the effective coupling
constant $|G_{\pi^0}(m^2_X)|$:\begin{eqnarray}\label{Eq6}
\Gamma(X(3872)\to\pi^0\chi_{c1}(1P);m^2_X)=(0.04\pm0.02)\
\mbox{MeV}, \qquad |G_{\pi^0}(m^2_X)|=(0.216\pm0.054)\
\mbox{GeV}^{-1}.\end{eqnarray}


\section{Loop mechanism of \boldmath{$X(3872)\to\pi^0\chi_{\lowercase{c}1}(1P)$}}

\begin{figure}  
\begin{center}\includegraphics[width=12cm]{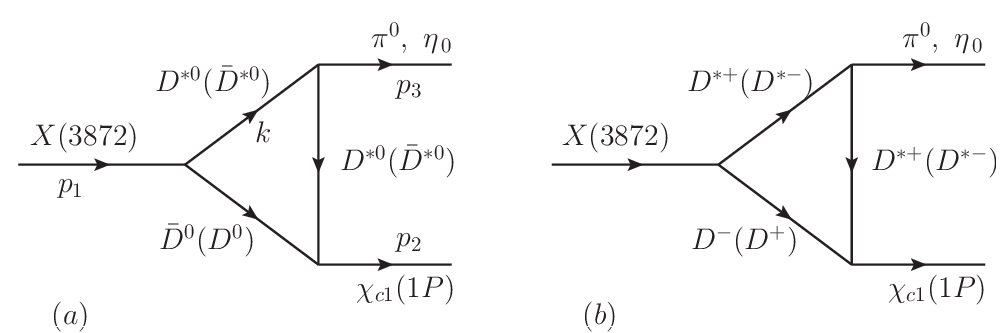}
\caption{\label{Fig1} The model of triangle loop diagrams for the
transitions $X(3872)\to(D\bar D^*+ \bar DD^*)\to(\pi^0, \,\eta_0)
\chi_{c1}(1P)$. }\end{center}\end{figure}

Let us consider the simplest model of triangle loop diagrams for the
amplitude $\mathcal{M}_{\pi^0} (s)$ introduced in Eq. (\ref{Eq1}).
It is graphically depicted in Fig. \ref{Fig1}. The specific
structure of the vertices in these diagrams is determined with use
of the effective Lagrangian,
\begin{eqnarray}\label{Eq7}
\mathcal{L}=ig_{XD\bar D^*}X^\mu(D^\dag D^*_\mu-DD^{*\dag}_\mu)+
ig_{\chi_{c1}D\bar D^*}\chi_{c1}^\mu(D^\dag D^*_\mu-DD^{*\dag}_\mu)+
g_{D^{*0}D^{*0}\pi^0}\varepsilon^{\mu\nu\lambda\kappa}\partial_\mu
D^*_\nu(\hat{\vec{\tau}}\vec{\pi}+\eta_0)\partial_\lambda D^{*\dag
}_\kappa,\end{eqnarray} where $D$, $D^\dag$, $D^*$, and $D^{*\dag}$
are the charm meson isodoublets, $\hat{\vec{\tau}}= (\hat{\tau}_1,
\hat{\tau}_2,\hat{\tau}_3)$ are the Pauli matrices, $\vec{\pi}=(
\pi_1,\pi_2,\pi_3)$ is the isotopic triplet of $\pi$ mesons, the
$\pi_3=\pi^0$ state has the quark structure $(u\bar u-d\bar
d)/\sqrt{2}$, and $\eta_0$ denotes isosinglet pseudoscalar state
with the quark structure $(u\bar u+d\bar d)/\sqrt{2}$. The amplitude
of the virtual $\eta_0$ state production, $\mathcal{M}_{\eta_0
}(s)$, will be useful to us in the following. For the coupling
constants indicated in Eq. (\ref{Eq7}), we introduce short
notations: $g_{XD\bar D^*}=g_X$, $g_{\chi_{c1}D\bar D^*}=g_{
\chi_{c1}}$, and $g_{D^{*0}D^{*0}\pi^0}=g_{\pi^0}$.
In accordance with Fig. \ref{Fig1}, we represent the amplitudes
$\mathcal{M}_{\pi^0}(s)$ and $\mathcal{M}_{\eta_0}(s)$ in the
following form:
\begin{eqnarray}\label{Eq8}
\mathcal{M}_{\pi^0}(s)=\frac{g_Xg_{\chi_{c1}}g_{\pi^0}}{16\pi}
\varepsilon^{\mu\nu\lambda\kappa}\epsilon^X_\mu(p_1)\epsilon^{
*\chi_{c1}}_\nu(p_2)[2C^n_\lambda(s)-2C^c_\lambda(s)]p_{3\kappa},
\end{eqnarray}
\begin{eqnarray}\label{Eq8a}
\mathcal{M}_{\eta_0}(s)=\frac{g_Xg_{\chi_{c1}}g_{\eta_0}}{16\pi}
\varepsilon^{\mu\nu\lambda\kappa}\epsilon^X_\mu(p_1)\epsilon^{
*\chi_{c1}}_\nu(p_2)[2C^n_\lambda(s)+2C^c_\lambda(s)]p_{3\kappa},
\end{eqnarray} where $g_{\eta_0}=g_{\pi^0}$,
the amplitudes $C^n_\lambda(s)$ and $C^c_\lambda(s)$ correspond to
the diagrams with neutral and charged particles in the loops,
respectively, and the factor 2 in front of them takes into account
that for each type of particles there are two such diagrams. The
amplitudes $C^n_\lambda(s)$ and $C^c_\lambda(s)$ are converged
separately and have the form,
\begin{eqnarray}\label{Eq9}
C^n_\lambda(s)=\frac{i}{\pi^3}\int\frac{k_\lambda\,d^4k}{(k^2-
m^2_{D^{*0}}+i\varepsilon)((p_1-k)^2-m^2_{\bar D^0}+i \varepsilon)(
(k-p_3)^2-m^2_{D^{*0}}+i\varepsilon)}=p_{1\lambda}C^n_{11}
(s)+p_{3\lambda}C^n_{12}(s),
\end{eqnarray}
\begin{eqnarray}\label{Eq10}
C^c_\lambda(s)=\frac{i}{\pi^3}\int\frac{k_\lambda\,d^4k}{(k^2-
m^2_{D^{*+}}+i\varepsilon)((p_1-k)^2-m^2_{D^-}+i \varepsilon)(
(k-p_3)^2-m^2_{D^{*+}}+i\varepsilon)}=p_{1\lambda}C^c_{11}
(s)+p_{3\lambda}C^c_{12}(s).
\end{eqnarray}
Substitution of Eqs. (\ref{Eq9}) and (\ref{Eq10}) into Eq.
(\ref{Eq8}) and comparison the result with Eq. (\ref{Eq1}) give [the
functions $C^n_{12}(s)$ and $C^c_{12}(s)$ do not contribute]
\begin{eqnarray}\label{Eq11}
\mathcal{M}_{\pi^0}(s)=-\frac{g_Xg_{\chi_{c1}}g_{\pi^0}}{16\pi}
\varepsilon^{\mu\nu\lambda\kappa}\epsilon^X_\mu(p_1)\epsilon^{
*\chi_{c1}}_\nu(p_2)p_{1\lambda}p_{2\kappa}[2C^n_{11}(s)-2C^c_{11}(s)],
\end{eqnarray}
\begin{eqnarray}\label{Eq12}
G_{\pi^0}(s)=-\frac{g_Xg_{\chi_{c1}}g_{\pi^0}}{16\pi}
\left[2C^n_{11}(s)-2C^c_{11}(s)\right].
\end{eqnarray}
The representation of invariant amplitudes $C^n_{11}(s)$ and
$C^c_{11}(s)$ via dilogarithms is well known
\cite{{tHV79,PV79,Ku87,Den07}}. However, it will be convenient for
us to calculate them using the dispersion method. To do this, we
shall first find their imaginary parts. They are determined by the
contributions of real intermediate states, i.e., contributions in
which both charmed mesons outgoing from the vertex of the $X(3872)$
decay are on the mass shell. Applying the Kutkosky rule \cite{Cu60}
to the amplitude $C^n_\lambda(s)$ [see diagram $(a)$ in Fig.
\ref{Fig1}], we find
\begin{eqnarray}\label{Eq13}
\mbox{Im}C^n_\lambda(s)=\frac{-|\vec{k}|}{2\pi\sqrt{s}}\int\frac{
k_\lambda\,d\cos\theta d\varphi}{(k-p_3)^2-m^2_{D^{*0}}}= \frac{-
|\vec{k}|}{2\pi \sqrt{s}}\int\frac{k_\lambda\,d\cos\theta d\varphi}{
m^2_{\pi^0}- 2k_0p_{30}+2|\vec{k}||\vec{p}_3|\cos\theta}=p_{1
\lambda}\mbox{Im} C^n_{11}(s)+p_{3\lambda}\mbox{Im}C^n_{12}(s),
\end{eqnarray} where $k_\lambda$ are the components of the four-momentum
$k=(k_0,\vec{k})$ of the intermediate $D^{*0}$ meson [outgoing from
the vertex of the $X(3872)$ decay] on its mass shell in the rest
frame of $X(3872)$, the polar angle $\theta$ and the azimuthal angle
$\varphi$ determine the direction of the vector $\vec{k}$ in the
reference frame with the $z$ axis directed along the momentum
$\vec{p}_3$; $k_0=(s+m^2_{D^{*0}}-m^2_{D^0})/(2\sqrt{s})$,
$|\vec{k}|=\sqrt{s^2-2s(m^2_{D^{*0}}+m^2_{D^0})+(m^2_{D^{*0}}-
m^2_{D^0})^2}/(2\sqrt{s})$, and
$p_{30}=(s+m^2_{\pi^0}-m^2_{\chi_{c1}})/(2\sqrt{s})$. After
calculating the scalar products $p^\lambda_1 \mbox{Im}C^n_\lambda
(s)$ and $p^\lambda_3\mbox{Im}C^n_\lambda(s)$, we get
\begin{eqnarray}\label{Eq14}
\mbox{Im}C^n_{11}(s)=\frac{1}{s|\vec{p}_3|^2}\left[|\vec{k}|p_{30}-
\left(k_0-\frac{1}{2}p_{30}\right)\frac{m^2_{\pi^0}}{2|\vec{p}_3|}\ln
\left(\frac{m^2_{D^{*0}}-t_-}{m^2_{D^{*0}}-t_+}\right)\right],
\end{eqnarray}
where $t_\pm=m^2_{D^{*0}}+m^2_{\pi^0}-2k_0p_{30}\pm2|\vec{k}||\vec{p
}_3|$ are the boundary values of the variable $t=(k-p_3)^2$ at
$\cos\theta=\pm1$.
\begin{figure}  
\begin{center}\includegraphics[width=14cm]{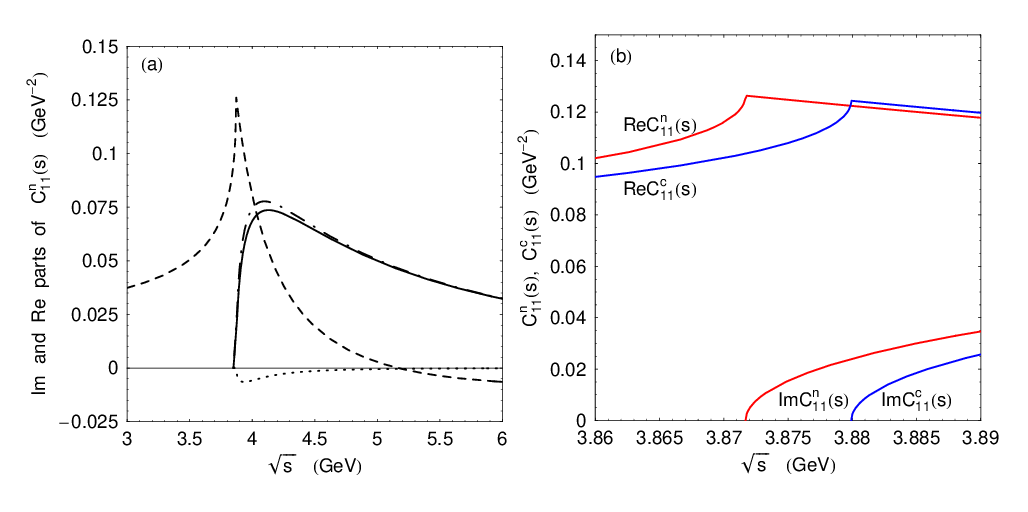}
\caption{\label{Fig2} (a) The solid and dashed curves show the
imaginary and real parts of the amplitude $C^n_{11}(s)$ constructed
using Eqs. (\ref{Eq14}) and (\ref{Eq15}), respectively, in a wide
region of $\sqrt{s}$. The dash-dotted curve shows the contribution
to $\mbox{Im}C^n_{11}(s)$ from the first (dominant) term in Eq.
(\ref{Eq14}) and the dotted curve shows the contribution from the
term containing a logarithm. (b) The imaginary and real parts of the
amplitudes $C^n_{11}(s)$ and $C^c_{11}(s)$ in the region of the
$D\bar D^*$ thresholds. }\end{center}\end{figure}
For $\sqrt{s}\gg4$ GeV, $\mbox{Im}C^n_{11}(s)\sim1/s$. We determine
the real part of the amplitude $C^n_{11}(s)$ numerically from the
dispersion relation,
\begin{eqnarray}\label{Eq15}
C^n_{11}(s)=\frac{1}{\pi}\int\limits^{\infty}_{s_n}\frac{
\mbox{Im}C^n_{11}(s')}{s'-s-i\varepsilon}ds',
\end{eqnarray}
where $s_n=(m_{D^0}+m_{\bar D^{*0}})^2$. Figure \ref{Fig2}(a) shows
the result of calculating the imaginary and real parts of the
amplitude $C^n_{11}(s)$ using Eqs. (\ref{Eq14}) and (\ref {Eq15}) in
a wide region of $\sqrt{s}$. Of course, we will ultimately be
interested a very narrow energy region near the $D^0\bar D^{*0}$
threshold where the $X(3872)$ object is located. The amplitude
$C^c_{11}(s)$ is calculated in exactly the same way. In the region
of the $D\bar D^*$ thresholds, the imaginary and real parts of the
amplitudes $C^n_{11}(s)$ and $C^c_{11}(s)$ are shown in Fig.
\ref{Fig2}(b), and the modulus and imaginary part of the difference
$2C^n_{11}(s)-2C^c_{11}(s)$ are shown in Fig. \ref{Fig3}(a). The $s$
dependence of the function $2C^n_{11}(s)-2 C^c_{11}(s)$ in this
region is well approximated by the difference between the rapidly
changing threshold factors $\rho^n(s)$ and $\rho^c(s )$ [see the
dotted curve in Fig. \ref{Fig3}(a) as an example]:
\begin{eqnarray}\label{Eq16}
2C^n_{11}(s)-2C^c_{11}(s)\simeq i[\rho^n(s)-\rho^c(s)]\times(0.692
\,\mbox{GeV}^{-2}),\end{eqnarray} where $\rho^n(s)=\sqrt{1-(m_{D^0}+
m_{\bar D^{*0}})^2/s}$ and $\rho^c(s)=\sqrt{1-(m_{D^+}+m_{D^{
*-}})^2/s}$ for $\sqrt{s}$ above the corresponding threshold,
and below one $\rho^n(s)\to i|\rho^n(s)|$ and $\rho^c(s)\to i|\rho^c
(s)|$. Note that at the $D^0\bar D^{*0}$ threshold
$2C^n_{11}((m_{D^0}+ m_{D^{*0}})^2)-2C^c_{11}((m_{D^0}+m_{D^{*0}}
)^2)\simeq|\rho^c((m_{ D^0}+m_{D^{*0}})^2)|\times(0.692\,
\mbox{GeV}^{-2})$; i.e., as a result of compensation, this
difference is determined by the remainder of the contribution of
charged intermediate states $D^+D^{*-}+D^- D^{*+}$. For $\sqrt{s}$
between the $D\bar D^*$ thresholds, we have
\begin{eqnarray}\label{Eq17}
|\rho^n(s)-\rho^c(s)|\simeq\sqrt{\frac{2(m_{D^+}+m_{D^{
*-}}-m_{D^0}-m_{\bar D^{*0}})}{m_{D^0}+m_{\bar D^{*0}}}}
\simeq0.0652.\end{eqnarray} Since the $X(3872)$ resonance is located
almost at the threshold of the $D^0\bar D^{*0}$ channel [see Fig.
\ref{Fig3}(a)], then the amplitude of the isospin-violating decay
$X(3872)\to\pi^0\chi_{c1} (1P)$, that is due to the considered loop
mechanism, turns out  to be proportional to $\sqrt{m_d-m_u}$ [see
Eqs. (\ref{Eq16}) and (\ref{Eq17})], rather than to $m_d-m_u$
[similar to the threshold effect of the $a_0(980)-f_0(980)$ mixing
\cite{AS79,AS19a}].
\begin{figure}  
\begin{center}\includegraphics[width=14cm]{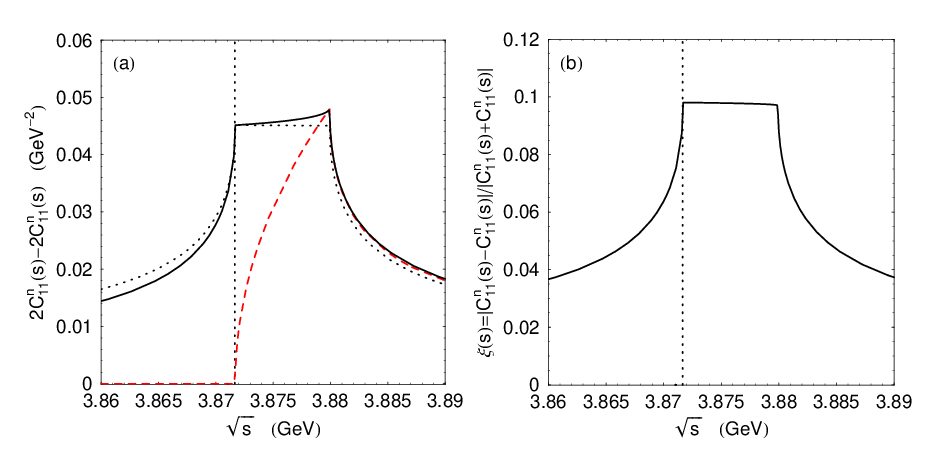}
\caption{\label{Fig3} (a) The solid and dashed curves show the
magnitude and imaginary part of the amplitude
$2C^n_{11}(s)-2C^c_{11}(s)$ from Eq. (\ref{Eq11}); the dotted curve
corresponds to the approximation of $|2C^n_{11}(s)-2C^c_{11}(s)|$
using Eq. (\ref{Eq16}). (b) The energy-dependent isospin violation
parameter $\xi(s)=|C^n_{ 11}(s)-C^c_{11}(s)|/|C^n_{11}(s)+C^c_{11}
(s)|$. The vertical dashed lines in (a) and (b) mark the position of
the $X(3872)$ resonance.}\end{center}\end{figure}

As a dimensionless parameter characterizing the scale of isospin
violation, it is natural to take the ratio of the production
amplitudes of the $\pi^0$ and $\eta_0$ states [see diagrams in Fig.
\ref{Fig1} and Eqs. (\ref{Eq8}) and (\ref{Eq8a})], i.e., the
quantity
\begin{eqnarray}\label{Eq18}
\xi(s)=\frac{|C^n_{11}(s)-C^c_{11}(s)|}{|C^n_{11}(s)+C^c_{11}
(s)|}.\end{eqnarray} The energy dependence of the parameter $\xi(s)$
is shown in Fig. \ref{Fig3}(b). At $\sqrt{s}=m_{X}=3871.65$ MeV
\cite{PDG23}, we have
\begin{eqnarray}\label{Eq19}
\xi=\xi(m^2_{X})\simeq0.916.\end{eqnarray} As we will see in the
next section, this value is a lower limit for $\xi$ in the
considered model. If the above estimate of $\xi$ being the relative
quantity can be rated as sufficiently reasonable, then to estimates
of the absolute values of the strong interaction amplitudes $C^n_{1
1}(s)$ and $C^c_{11}(s)$ [see Figs. \ref{Fig2} and \ref{Fig3}(a)],
we should treat with the extreme caution. Here, we mean the need to
take into account the influence of the form factor on these
amplitudes in order to obtain physically more meaningful estimates
for them. We discuss this issue below.


\section{Estimate of the amplitude \boldmath{$X(3872)\to\pi^0\chi_{\lowercase{c}1}(1P)$}}


In order to take into account to some extent the internal structure
and the off-mass-shell effect for the $D^*$ meson, by which there is
the exchange between the intermediate $D(\bar D)$ and $\bar D^*(D^*)
$ mesons in the triangle loops (see Fig. \ref{Fig1}), it is
necessary to introduce the form factor into each vertex of the $D^*$
exchange,
\begin{eqnarray}\label{Eq20}
\mathcal{F}(q^2,m^2_{D^*})=\frac{\Lambda^2-m^2_{D^*}}{\Lambda^2-q^2},
\end{eqnarray}where $\Lambda$ is the cutoff parameter, $m_{D^*}$ and
$q$ are the mass and four-momentum of the exchanged $D^*$ meson,
respectively. Such a type of the monopole form factor was first used
in \cite{Lo94,Go96} to calculate triangle loops when describing the
annihilation process at rest $p\bar p\to\pi\phi$, introduced into
use \cite{Co02,Co04} for estimating rescattering effects in $B^-\to
K^-\chi_{c0}$, $B^-\to K^-h_c$ decays, discussed in detail in
calculations of final state interactions in various hadronic $B$
meson decay channels \cite{Ch05}, and is now widely used in
describing loop mechanisms of heavy quarkonium decays; see, for
example, \cite{Li07,Me07,Li12,Wu21,Ba22,Wa22} and references herein.
The standard form of the parameter $\Lambda$ is\cite{Ch05}
$\Lambda=m_{D^*}+\alpha\Lambda_{\scriptsize\mbox{QCD}}$, where
$\Lambda_{\scriptsize\mbox{QCD}}=220$ MeV and {\it a priori} unknown
value of $\alpha$ is found from fitting the data. Let us rewrite Eq.
(\ref{Eq20}) as follows: $\mathcal{F}(q^2,m^2_{ D^*})=\frac{1}{1+
(m^2_{D^*}-q^2)/(\Lambda^2- m^2_{D^*})}$. From here, it is clear
that the parameter $1/(\Lambda^2-m^2_{D^*})$ determines the rate of
change of the form factor when the $D^*$ meson leaves the mass
shell.

\begin{figure}  
\begin{center}\includegraphics[width=14cm]{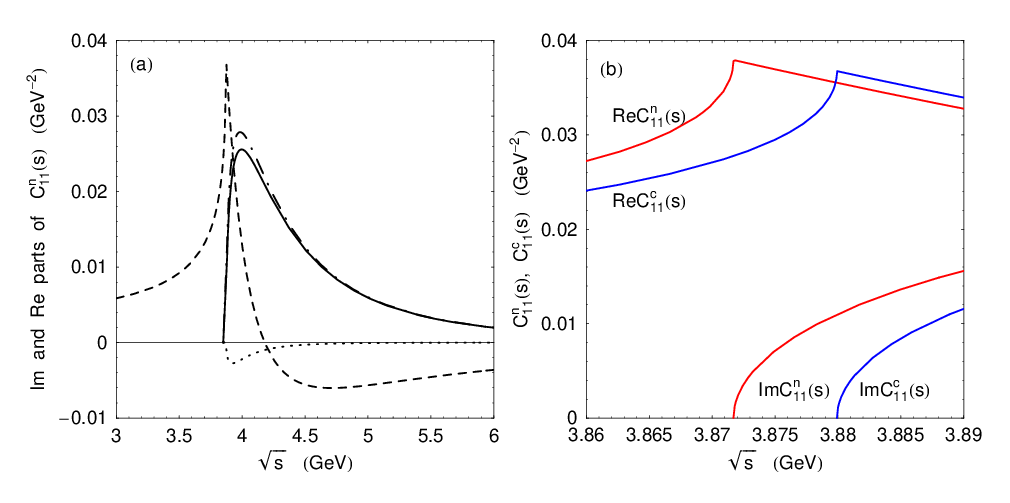}
\caption{\label{Fig4} (a) The solid and dashed curves show the
imaginary and real parts of the amplitude $C^n_{11}(s)$,
respectively, constructed in a wide region of $\sqrt{s}$ taking into
account the form factor, see Eq. (\ref{Eq20}), at $\alpha=2.878$
($\Lambda \simeq2.64$ GeV) according to  Eqs. (\ref{Eq15}) and
(\ref{Eq21})--(\ref{Eq23}). The dash-dotted curve shows the
contribution to $\mbox{Im}C^n_{11}(s)$  from the first term in Eq.
(\ref{Eq14}) modified according to Eq. (\ref{Eq22}), and the dotted
curve is from the term containing logarithm modified according to
Eq. (\ref{Eq23}). (b) Imaginary and real parts of the amplitudes
$C^n_{11}(s)$ and $C^c_{11}(s)$ in the region of the $D\bar D^*$
thresholds taking into account the form factors at $\alpha=2.878.$
}\end{center}\end{figure}

Let us now write the expression for $\mbox{Im}C^n_\lambda(s)$ [see.
Eq. (\ref{Eq13})] taking into account the form factor,
\begin{eqnarray}\label{Eq21}
\mbox{Im}C^n_\lambda(s)=p_{1\lambda}\mbox{Im}
C^n_{11}(s)+p_{3\lambda}\mbox{Im}C^n_{12}(s)=\frac{-|\vec{k}|}{2\pi
\sqrt{s}}\int\frac{ \mathcal{F}^2(q^2,m^2_{D^{*0}}) \,k_\lambda\,d
\cos\theta d\varphi}{(k-p_3 )^2-m^2_{D^{*0}}},
\end{eqnarray} where $q^2=(k-p_3 )^2$, and carry out the
corresponding calculations. As a result, the first term in Eq.
(\ref{Eq14}) is multiplied by
\begin{eqnarray}\label{Eq22}
\frac{(\Lambda^2-m^2_{D^{*0}})^2}{(\Lambda^2-t_+)(\Lambda^2-t_-)}
\end{eqnarray} and $\ln\left[(m^2_{D^{*0}}-t_-)/( m^2_{
D^{*0}}-t_+)\right]$ is replaced by
\begin{eqnarray}\label{Eq23}
\ln\left[\frac{(m^2_{D^{*0}}-t_-)(\Lambda^2-t_+)}{(m^2_{D^{*0}}-t_+)
(\Lambda^2-t_-)}\right]-\frac{(\Lambda^2-m^2_{D^{*0}})(t_+-t_-)}{(
\Lambda^2-t_+)(\Lambda^2-t_-)}.
\end{eqnarray}
Note that in the case under consideration, the virtuality of the
$D^{*0}$-meson, i.e., $(m^2_{D^{*0}}-q^2)$ turns out to be greater
than 1.373 GeV$^2$. At $\sqrt{s}\gg4$ GeV, the amplitude $\mbox{Im}
C^n_{11}(s)$ taking into account the form factor falls as $1/s^2$.
The real part of $C^n_{11}(s)$ is determined numerically from the
dispersion relation (\ref{Eq15}). The amplitude $C^c_{11}(s)$ taking
into account the form factor is calculated in exactly the same way.
Figure \ref{Fig4}(a) shows as an example the result of the
calculation of the imaginary and real parts of the amplitude
$C^n_{11}(s)$ taking into account the form factor (\ref{Eq20}) at
$\alpha=2.878$ ($\Lambda\simeq2.64$ GeV) in a wide region of $
\sqrt{s}$. In the region of the $D\bar D^*$ thresholds, the
imaginary and real parts of the amplitudes $C^n_{11}(s)$ and
$C^c_{11}(s)$ taking into account form factors at $\alpha=2.878$ are
shown in Fig. \ref{Fig4}(b). Comparison of the curves in Fig.
\ref{Fig4}(b) with those in Fig. \ref{Fig2}(b), which correspond to
$\mathcal{F}^2(q^2,m^2_{ D^*})\equiv1$ (i.e., $\alpha=\infty$),
shows that the form factor with $\alpha=2.878$ reduces the
amplitudes near the $D\bar D^*$ thresholds by approximately 3.5
times.

\begin{figure}  
\begin{center}\includegraphics[width=14cm]{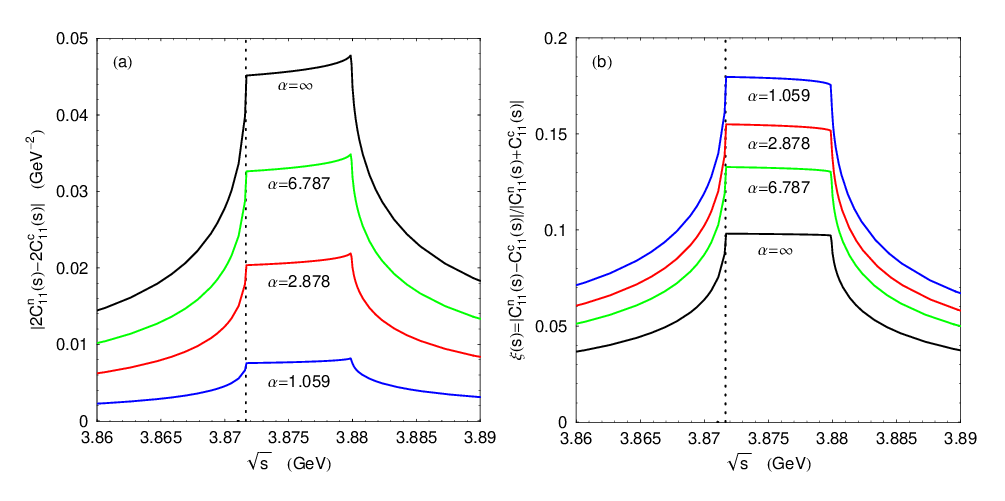}
\caption{\label{Fig5} (a) Modulus of the amplitude $2C^n_{11}(s)-
2C^c_{11}(s)$ for several values of the parameter $\alpha$. (b) The
energy-dependent isospin violation parameter $\xi(s)=|C^n_{11}(s)-
C^c_{11}(s)|/|C^n_{11}(s)+C ^c_{11} (s)|$ for the same values of
$\alpha$. The vertical dotted lines in (a) and (b) mark the position
of the $X(3872)$ resonance.}\end{center}
\end{figure}
\begin{figure}  
\begin{center}\includegraphics[width=14cm]{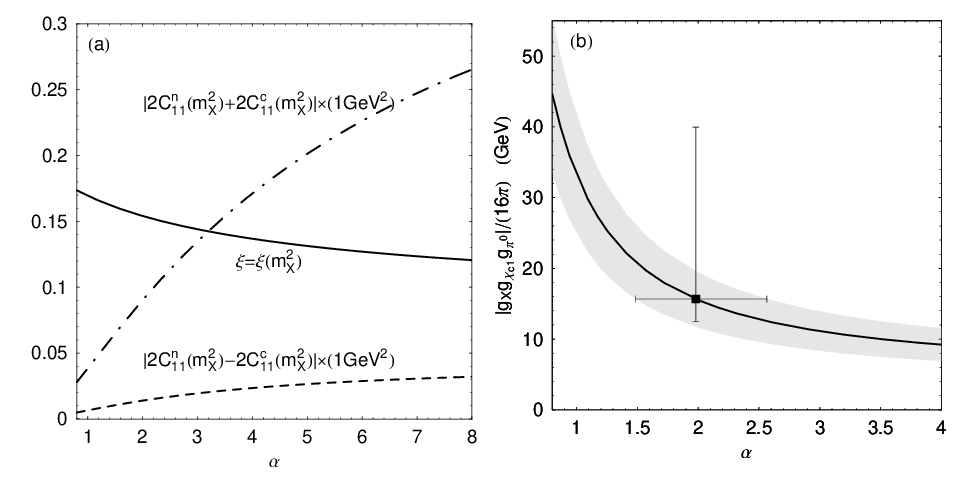}
\caption{\label{Fig6} (a) The isospin violation parameter
$\xi=\xi(m^2_X)$ and dimensionless amplitudes $|2C^n_{ 11}(m^2_X
)-2C^c_{11}(m^2_X)|\times(1\,\mbox{GeV}^2)$ and $|2C^n_{ 11}(m^2_X
)+2C^c_{11}(m^2_X)|\times(1\,\mbox{GeV}^2)$ as functions of $\alpha$
[for $\alpha=\infty$ ($\Lambda=\infty$), i.e., when $\mathcal{F}^2
(q^2, m^2_{D^*})\equiv1$, the asymptotics of these quantities are
0.916, 0.0419, and 0.457, respectively]. (b) The shaded band shows
the dependence on $\alpha$ of the values of the right-hand side of
Eq. (\ref{Eq24}) lying within the uncertainty of the quantity
$|G_{\pi^0}(m^2_X)|$; the solid curve inside the band corresponds to
the central value of $|G_{\pi^0}(m^2_X)|$. The dot with vertical
error bars shows the estimate presented in Eq. (\ref{Eq27}) for the
left side of Eq. (\ref{Eq24}); the horizontal segment of the
straight line marks the interval of $\alpha$ values at which the Eq.
(\ref{Eq24}) is consistent. }\end{center}\end{figure}

Let us now trace with the help of Figs. \ref{Fig5} and \ref{Fig6}(a)
for the influence of the form factor on the modulus of the amplitude
difference $2C^n_{11}(s)-2C^c_{11}(s)$, the parameter $\xi(s)$ and
its particular value $\xi=\xi(m^2_X )$ [see Eqs. (\ref{Eq18}) and
(\ref{Eq19})]. As can be seen from the examples shown in Fig.
\ref{Fig5}, $|2C^n_{ 11}(s)-2C^c_{11}(s)|$ and $\xi(s)$ have
opposite dependences on $\alpha$. With increasing suppression of the
$|2C^n_{11}(s)-2C^c_{11}(s)|$ amplitude by the form factor (i.e.,
with decreasing $\alpha$), $\xi(s)$ increases. For $\sqrt{s}=m_X$,
$|2C^n_{11}(m^2_X)-2C^c_{11}(m^2_X)|$ and $\xi=\xi(m^2_X) $ as
functions of $\alpha$ are shown in Fig. \ref{Fig6}(a). This figure
also explains why there is an increase in isospin violation, i.e.,
increasing the parameter $\xi=\xi(m^2_X)$, with decreasing $\alpha$.
This behavior of $\xi$ is due to different suppression rate of the
amplitudes $|2C^n_{11}(m^2_X)-2C^c_{11} (m^2_X)|$ and $|2C^n_{11}
(m^2_X)+2C^c_{11}( m^2_X)|$ with decreasing $\alpha$ (or $\Lambda$)
in the form factor; see the dashed and dash-dotted curves in Fig.
\ref{Fig6}(a).

Now we are ready to estimate the absolute value of the
$X(3872)\to\pi^0\chi_{c1}(1P)$ decay amplitude. First of all, we
indicate those values of the product of coupling constants
$|g_Xg_{\chi_{c1 }}g_{\pi^0}|/(16\pi)$ for which the considered
model can be consistent with available data. Using Eqs. (\ref{Eq6})
and (\ref{Eq12}), we write
\begin{eqnarray}\label{Eq24}
\frac{|g_Xg_{\chi_{c1}}g_{\pi^0}|}{16\pi}=\frac{|G_{\pi^0}(m^2_X)|}{
|2C^n_{11}(m^2_X)-2C^c_{11}(m^2_X)|}=\frac{(0.216\pm0.054)\
\mbox{GeV}^{-1}}{|2C^n_{11}(m^2_X)-2C^c_{11}(m^2_X)|}.
\end{eqnarray}
From Eq. (\ref{Eq24}), it follows that the suitable values of
$|g_Xg_{\chi_{c1}}g_{\pi^0}|/(16\pi)$ (for reasonable values of
$\alpha$) lie in the shaded band shown in Fig. \ref{Fig6}(b). The
band is due to the uncertainty in the value of $|G_{\pi^0}(m^2_X)|$.
The solid curve inside the band corresponds to the central value of
$|G_{\pi^0}(m^2_X)|$. In the absence of the form factor, i.e., for
$\alpha=\infty$, for $|g_Xg_{\chi_{c1}}g_{\pi^0}|/(16\pi)$ is
predicted the range of values from 3.87 to 6.45 GeV. If $|g_Xg_{
\chi_{c1}}g_{\pi^0}|/ (16\pi)<3.87$ GeV, then the model is
unsatisfactory. Sources of information about the constants $g_X$,
$g_{\chi_{c1}}$ and $g_{\pi^0}$, which determine the left side of
Eq. (\ref{Eq24}), are the data on the $X(3872)\to(D^0\bar
D^{*0}+\bar D^ 0D^{*0})\to D^0\bar D^0\pi^0$ and $X(3872)\to\pi^+
\pi^-J/\psi$ decays and theoretical considerations. An approximate
value of $g_X\equiv g_{XD\bar D^*}\equiv g_{XD^0\bar D^{*0}}$ [see.
Eq. (\ref{Eq7})] we will take from the processing of the data on the
$X(3872)$ decays obtained by the Belle \cite{Aus10} (for processing
see Ref. \cite{AR14}), LHCb \cite{Aai20}, Belle \cite{Hir23,Tan23},
and BESIII \cite{Abl23} Collaborations. The coupling constant $g_X$
in Ref. \cite{AR14} was denoted as $g_A$. Let us note that the
fitted parameter used in Refs. \cite{Aai20,Hir23,Tan23,Abl23} was
the coupling constant $g$, which is related to $g_X$ by the relation
$g=g^2_X/(4\pi m^2_X)$. Information about the values of $g$ and
$g_X$ and their statistical errors are collected in Table I. The
lower limits for $g$ were also obtained in Refs. \cite{Hir23,Tan23}:
$g>0.075$ ($g_X>3.76$ GeV) and $g>0.094$ ($g_X>4.21$ GeV) at 95\%
and 90\% confidence level, respectively. Some difficulties with
determining the value of $g$ (partly associated with limited
statistics) and the estimates of systematic uncertainties are
discussed in detail in Refs. \cite{Aai20,Hir23, Tan23,Abl23}. Here,
we only note that the sensitivity of $g$ to the mass of $X(3872)$
(caused by its proximity to the $D^0\bar D^{*0}$ threshold) and weak
dependence of the $X (3872)$ line shape in the $D^0\bar D^0\pi^0$
channel on $g$ at large $g$ generate significant positive
uncertainties in this constant in the fits. In our opinion, a large
positive error in $g$ should not be given any decisive significance
compared to the central value of $g$. New experiments with high
statistics should clarify the situation. For our purposes, we will
use the average value of $g_X=\left(5.81^{+8.97 }_{-0.82}\right)$
GeV found from the data in Table I.
\begin{center}\begin{table}  [!ht]
\caption{Information about the $X(3872)$ coupling to the $D^0\bar
D^{*0}$ system.}
\begin{tabular}{ c c c c c } \hline\hline
  \ \,Data analysis\ \ &\ \ AR \cite{AR14}\ \ &\ \ LHCb \cite{Aai20}\ \ &\ \ Belle \cite{Hir23,Tan23}\ \ &\ \ BESIII \cite{Abl23}\ \ \\
  \hline
  $g$ & $0.181^{+0.647}_{-0.127}$ & $0.108\pm0.003$  & $0.29^{+2.69}_{-0.15}$  &  $0.16\pm0.10$\ \ \vspace*{0.1cm}\\
  $g_X$ (GeV) & $5.85^{+10.42}_{-2.04}$ & $4.51\pm0.06$  & $7.39^{+34.28}_{-1.91}$ & $5.49\pm1.72$\ \ \\
  \hline\hline\end{tabular}\end{table}\end{center}

To estimate the constants $g_{\chi_{c1}}\equiv g_{\chi_{c1}D\bar
D^*}$ and $g_{\pi^0}\equiv g_{D^{*0}D^ {*0}\pi^0}$ [see Eq.
(\ref{Eq7})] we use the results obtained in Refs. \cite{Wu21,
Me07,Co02,Co04, Ch05,Gu11} in the framework of the heavy quark
effective theory:
\begin{eqnarray}\label{Eq25} 
g_{\chi_{c1}D\bar D^*}=2\sqrt{2}g_1\sqrt{m_Dm_{D^*}m_{\chi_{c1}}},
\quad g_1=-\frac{\sqrt{m_{\chi_{c0}}/3}}{f_{\chi_{c0}}}, \quad
f_{\chi_{c0}}=(510\pm40)\ \mbox{MeV}, \\
g_{D^{*0}D^{*0}\pi^0}= \frac{g_{D^{*0}D^0\pi^0}}{\sqrt{m_Dm_{D^*}}}
=\frac{\sqrt{2}g}{f_\pi}, \quad f_\pi=132\ \mbox{MeV}, \quad g=0.59
\pm0.07.\end{eqnarray}  Thus we have $g_{\chi_{c1}}\equiv g_{\chi_{
c1}D \bar D^*}=(-21.45\pm1.68)$ GeV, $g_{\pi^0}\equiv
g_{D^{*0}D^{*0} \pi^0 }=(6.32\pm0.75)$ GeV$^{-1}$, and
\begin{eqnarray}\label{Eq27} \frac{|g_Xg_{\chi_{c1}}g_{\pi^0}|}{16\pi}
=\left(15.67^{+24.29}_{-3.14}\right)\ \mbox{GeV}. \end{eqnarray} The
value (\ref{Eq27}) is shown in Fig. \ref{Fig6}(b) in the form of a
dot with vertical error bars. Agreement with the data on the
amplitude $|G_{\pi^0}(m^2_X)|$ [see Eqs. (\ref{Eq6}) and
(\ref{Eq24})] is achieved when this point falls inside the shaded
band. This occurs in the $\alpha$ interval from 1.487 to 2.565
marked in Fig. \ref{Fig6}(b) by a segment of a horizontal straight
line. At $\alpha=1.98$, the central values of the left and right
sides of Eq. (\ref{Eq24}) coincide. In the indicated interval of
$\alpha$, the average value of the isospin violation parameter $\xi$
is of about 0.15; see Fig. \ref{Fig6}(a).
\begin{figure}  
\begin{center}\includegraphics[width=7cm]{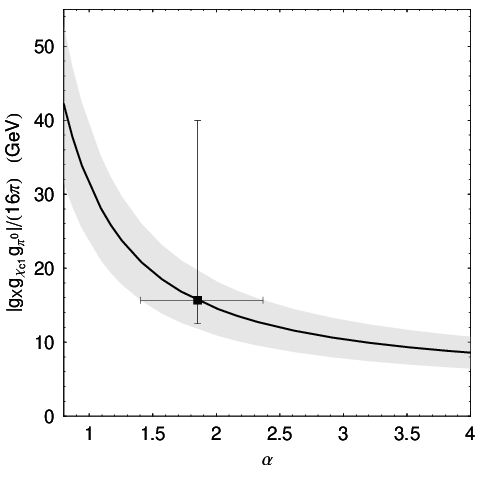}
\caption{\label{Fig7} The same as in Fig. \ref{Fig6}(b) but with
taking into account the $\pi^0-\eta$ mixing, see the text.}
\end{center}\end{figure}
For comparison, we point out that the isospin violation parameter
for the $\pi^0$ production mechanism due to the $\pi^0-\eta$ mixing
is an order of magnitude smaller \cite{Fe00}: $\Pi_{\pi^0\eta}/
(m^2_\eta-m^2_{\pi^0})\simeq0.014$, where $\Pi_{\pi^0\eta}$ is the
$\pi^0\leftrightarrow\eta$ transition amplitude having dimension of
a mass squared. Taking into account the mechanism of the $\pi^0-
\eta$ mixing and the relation $\eta_0=\eta\sin(\theta_i-\theta_p
)+\eta'\cos(\theta_i-\theta_p)$, where $\eta$ and $\eta'$ are the
physical states of the lightest pseudoscalar isoscalar mesons, Eq.
(\ref{Eq12}) takes the form,
\begin{eqnarray}\label{Eq28}
G_{\pi^0}(s)=-\frac{g_Xg_{\chi_{c1}}g_{\pi^0}}{16\pi}
\left[2C^n_{11}(s)-2C^c_{11}(s)+\sin(\theta_i-\theta_p)\frac{\Pi_{
\pi^0\eta}}{m^2_\eta- m^2_{\pi^0}}(2C^n_{11}(s)+2C^c_{11}(s))
\right].\end{eqnarray} Here $\theta_i=35.3^\circ$ is the so-called
``ideal'' mixing angle and $\theta_p=-11.3^\circ$ is the mixing
angle in the nonet of the light pseudoscalar mesons \cite{PDG23}.
The result of analyzing Eq. (\ref{Eq28}) is shown in Fig.
\ref{Fig7}. This result is similar to that based on Eq. (\ref{Eq24})
and shown in Fig. 6(b). Now the permissible values of $\alpha$ lie
in the range from 1.406 to 2.368, and the central value of $\alpha$
is equal to 1.853; i.e., changes in $\alpha$ turn out to be less
than 10\%. Note that the parameter $\alpha$ confirms its status as
an useful fitting parameter with expected fitted values of the order
of 1. Improving data accuracy on the width of the $X(3872)\to
\pi^0\chi_{c1}(1P)$ decay is one of the great demand and essential
task. Our conclusions from the present analysis are briefly
formulated in the next section.

\section{Conclusion}

Thus, we conclude that the considered model of triangle loops for
the decay amplitude $X(3872)\to\pi^0\chi_{c1}(1P)$ is generally in
reasonable agreement with the available data. Its distinctive
feature is the convergence of diagrams with neutral and charged
charmed mesons in the loops separately and without taking into
account the form factor.

The significant amplitude of the process $X(3872)\to\pi^0\chi_{c1}
(1P)$, which violates isospin, indicates the threshold nature of the
origin of this effect. Due to incomplete compensation of the
contributions of the $D^{*0}\bar D^0D^{*0}+c.c.$ and $D^{*+}D^-D^{
*+}+c.c.$ loops, caused by the differences in the masses
$m_{D^+}- m_{D^0}$ and $m_{D^{*+}}-m_{D^{*0}}$, the amplitude
$X(3872)\to\pi^0\chi_{c1}(1P)$ near the $D^{*0}\bar D^0$ threshold
turns out to be proportional to $\sqrt{m_d-m_u}$, and not $m_d-m_u$.
That is, the mechanism of the charmed meson loops manifests itself
at a qualitative level.

The product of the coupling constant $|g_Xg_{\chi_{c1}}g_{\pi^0}
|/(16\pi)$ and parameter $\alpha$ accumulate important information
about the interactions of the $X(3872)$, $\chi_{c1}(1P)$, $D$, $D^*
$, and $\pi$ mesons and determine the loop mechanism of the process
$X(3872)\to\pi^0\chi_{c1}(1P)$ in accordance with existing data.

Apart from the difference in the masses of neutral and charged
charmed mesons, any additional exotic sources of isospin violation
in $X(3872)\to\pi^0\chi_{c1}(1P)$ (such as a significant difference
between the coupling constants $g_{XD^0\bar D^{*0}}$ and $g_{XD^+
D^{*-}}$) are not required to interpret the data. This indirectly
confirms the isotopic neutrality of the $X(3872)$, which is
naturally realized for the $c\bar c$ state $\chi_{c1}(2P)$.

Increasing data accuracy about the $X(3872)$ in all directions [in
particular, on the $X(3872)\to\pi^0\chi_{c1}(1P)$ decay] will
certainly shed light on the mysterious nature of this extraordinary
state.

Here, it would also be appropriate to note the importance of modern
studies of the $X(3872)$ state in the molecular model. This model is
significantly has evolved and extended its predictions to a large
number of specific processes; see Refs. \cite{Sw04,Zh14,Me15,Wu21,
Wa22,Gu14,Wu23} and references herein. For example, recently in Ref.
\cite{Wu23}, using a molecular approach within the framework of the
triangle diagram model, the large experimentally observed violation
of the isospin symmetry in the $\mathcal{B}(B^+\to X(3872)K^+)
/\mathcal{B}(B^0\to X(3872)K^0)$ ratio was explained. In the
molecular model, the $X(3872)$ is formed by neutral $D^0\bar
D^{*0}+\bar D^0D^{*0}$ and charged $D^+D^{*-}+D ^- D^{*+}$ charmed
meson pairs. Verification in different processes of model
predictions based on the universality (i.e., independence from the
process) of the couplings of $X(3872)$ to its neutral and charged
constituents (the values of these couplings are different) seems to
be extremely important for the molecular scenario.\\


\begin{center} {\bf ACKNOWLEDGMENTS} \end{center}

The work was carried out within the framework of the state contract
of the Sobolev Institute of Mathematics, Project No. FWNF-2022-0021.



\begin{thebibliography}{99}
\bibitem{FN1} In Ref. \cite{Wu21}, the widths $\Gamma (X(387 2)\to
\pi^+\pi^-J/\psi)$ and $\Gamma(X(3872) \to\pi^+\pi^-\pi^0 J/ \psi)$
have been estimated via intermediate charmed meson loops with the
same effective coupling constant ($g_X $) of $X(3872)$ with its
components. In so doing, any additional normalization factors
$1/W_\rho$ and $1/W_\omega$ are not required in the definition of
these widths ($W_\rho\neq W_\omega$).
\end{thebibliography}
\end{document}